\newcommand{\beq}{\begin{equation}}
\newcommand{\eeq}{\end{equation}}
\newcommand{\bea}{\begin{eqnarray}}
\newcommand{\eea}{\end{eqnarray}}

\newcommand{\gsim}{\lower.7ex\hbox{$\;\stackrel{\textstyle>}{\sim}\;$}}
\newcommand{\lsim}{\lower.7ex\hbox{$\;\stackrel{\textstyle<}{\sim}\;$}}

\documentclass[aps,prd,nofootinbib,twocolumn,superscriptaddress,preprintnumbers,balancelastpage,longbibliography]{revtex4-1}
\pdfoutput=1
\usepackage{subfigure}
\usepackage{graphicx}
\usepackage{epstopdf}
\usepackage{mathrsfs}
\usepackage{amssymb}
\usepackage{verbatim}
\usepackage{color}
\usepackage{multirow}
\usepackage{amsmath}
\usepackage{hyperref}

\hypersetup{pdfstartview=FitV,colorlinks=true,linkcolor=blue,citecolor=red,filecolor=black,urlcolor=blue}

\def\stacksymbols #1#2#3#4{\def\theguybelow{#2}
    \def\vp{\lower#3pt}
    \def\sp{\baselineskip0pt\lineskip#4pt}
    \mathrel{\mathpalette\intermediary#1}}

\def\intermediary#1#2{\vp\vbox{\sp
     \everycr={}\tabskip0pt
     \halign{$\mathsurround0pt#1\hfil##\hfil$\crcr#2\crcr
              \theguybelow\crcr}}}

\def\trh{T_{\rm RH}}

\def\arh{a_{\rm RH}}
\def\aosc{a_{\rm osc}}

\def\be{\begin{equation}}
\def\ee{\end{equation}}
\def\bea{\begin{eqnarray}}
\def\eea{\end{eqnarray}}

\def\sp{\;\;\;,\;\;\;}

\def\arh{a_{\rm RH}}
\def\aosc{a_{\rm osc}}
\def\achi{a_{\chi}}
\def\aend{a_{\rm end}}

\def\lsim{\raise0.3ex\hbox{$\;<$\kern-0.75em\raise-1.1ex\hbox{$\sim\;$}}}
\def\gsim{\raise0.3ex\hbox{$\;>$\kern-0.75em\raise-1.1ex\hbox{$\sim\;$}}}

\def\inbar{\,\vrule height1.5ex width.4pt depth0pt}

\def\IC{\relax\hbox{$\inbar\kern-.3em{\rm C}$}}
\def\IQ{\relax\hbox{$\inbar\kern-.3em{\rm Q}$}}
\def\IR{\relax{\rm I\kern-.18em R}}
 \font\cmss=cmss10 \font\cmsss=cmss10 at 7pt
\def\IZ{\relax\ifmmode\mathchoice
 {\hbox{\cmss Z\kern-.4em Z}}{\hbox{\cmss Z\kern-.4em Z}}
 {\lower.9pt\hbox{\cmsss Z\kern-.4em Z}}
 {\lower1.2pt\hbox{\cmsss Z\kern-.4em Z}}\else{\cmss Z\kern-.4em Z}\fi}

\def\comment#1{}
\def\to{\rightarrow}

\def\u1x{U(1)_X}
\newcommand{\nc}{\newcommand}
\nc{\LL}{L}
\nc{\vv}{\tilde{v}}
\nc{\ccdot}{\!\cdot\!}
\nc{\gsm}{G_{SM}}
\nc{\vfive}{\mathbf{5}\oplus\mathbf{\overline{5}}}
\nc{\vten}{\mathbf{10}\oplus\mathbf{\overline{10}}}
\nc{\zhol}{Z^{\rm hol}}
\nc{\xfb}{\,{\rm fb}}

\begin{document}
\preprint{UMN--TH--4321/24}
\preprint{FTPI--MINN--24/13}

\title{Inflaton Production of Scalar Dark Matter through Fluctuations and Scattering}

\author{Gongjun Choi}
\affiliation{ 
 William I.~Fine Theoretical Physics Institute, 
       School of Physics and Astronomy,
            University of Minnesota, Minneapolis, MN 55455, USA
            }
\author{Marcos~A.~G.~Garcia}
\affiliation{Departamento de F\'isica Te\'orica, Instituto de F\'isica, Universidad Nacional Aut\'onoma de M\'exico, Ciudad de M\'exico C.P. 04510, Mexico}
\author{Wenqi Ke}
\affiliation{
 William I.~Fine Theoretical Physics Institute, 
       School of Physics and Astronomy,
            University of Minnesota, Minneapolis, MN 55455, USA
            }
\author{Yann Mambrini}
\affiliation{Sorbonne Universit\'e, CNRS, Laboratoire de Physique Th\'eorique et Hautes Energies, LPTHE, F-75005 Paris, France}
\author{Keith A. Olive}
\affiliation{ 
 William I.~Fine Theoretical Physics Institute, 
       School of Physics and Astronomy,
            University of Minnesota, Minneapolis, MN 55455, USA
            }
\author{Sarunas Verner}
\affiliation{Institute for Fundamental Theory, Physics Department, University of Florida, Gainesville, FL 32611, USA}

\begin{abstract} 
We study the effects on particle production of a Planck-suppressed coupling between the inflaton and a scalar dark matter candidate, $\chi$. In the absence of this coupling the dominant source for the relic density of $\chi$ is the long wavelength modes produced from the scalar field fluctuations during inflation. In this case, there are strong constraints on the mass of the scalar and the reheating temperature after inflation from the present-day relic density of $\chi$ (assuming $\chi$ is stable). When a coupling $\sigma \phi^2 \chi^2$ is introduced, with
$\sigma = {\tilde \sigma} m_\phi^2/ M_P^2 \sim 10^{-10} {\tilde \sigma}$, where $m_\phi$ is the inflaton mass, the allowed parameter space begins to open up considerably even for ${\tilde \sigma}$ as small as $\gtrsim 10^{-7}$. For ${\tilde \sigma} \gtrsim \frac{9}{16}$, particle production is dominated by the scattering of the inflaton condensate, either through single graviton exchange or the contact interaction between $\phi$ and $\chi$. In this regime, the range of allowed masses and reheating temperatures is maximal. For $0.004 < {\tilde \sigma} < 50$, constraints from isocurvature fluctuations are satisfied, and the production from parametric resonance can be neglected. 
\end{abstract}

\maketitle


\section{Introduction}
\label{sec:intro}

Quantum scalar fluctuations are an inevitable consequence of a de Sitter phase, including inflation \cite{fluc}. During inflation, 
these fluctuations grow linearly in time, reaching an asymptotic value $\langle \chi^2 \rangle \sim H_I^4/m_\chi^2$ for a massive
scalar field, $\chi$, with a mass (much) smaller
than the Hubble scale during inflation, $m_\chi \ll H_I$. For sufficiently light scalar fields, these fluctuations are very large, and the long wavelength contributions of these fluctuations obey the classical equations of motion \cite{longw}, making them indistinguishable from a homogeneous scalar field background. 

Although there are far too many cosmological consequences of these fluctuations to delineate here, we note their particular importance along (SUSY) flat directions \cite{gkm} and their role in Affleck-Dine Baryogenesis \cite{AD,Linde:1985gh,Campbell:1986qg,Enqvist:2003gh,Garcia:2013bha}, in avoiding the washout of density perturbations \cite{Graziani:1988bp,Enqvist:2003gh,Enqvist:2011pt}, and on reheating and thermalization \cite{Allahverdi:2005mz,Olive:2006uw}. All of these effects rely on the fact that the flat directions (or very light scalar fields) remain so during inflation. In general supergravity models, this is difficult \cite{Dine:1995uk} as Hubble scale masses are generally induced for all scalar fields. This is not the case for theories with a Heisenberg symmetry \cite{Binetruy:1987xj} which includes no-scale supergravity \cite{noscale}, where flat directions are maintained during inflation \cite{GMO}.

Of particular interest to us here are the consequences of the fluctuations of relatively light {\em stable} scalar fields, which can eventually play the role of cold dark matter \cite{Turner:1987vd,Peebles:1999fz,Enqvist:2014zqa,Nurmi:2015ema,Bertolami:2016ywc,Alonso-Alvarez:2018tus,Markkanen:2018gcw,Tenkanen:2019aij,Choi:2019mva,Cosme:2020nac,Lebedev:2022cic}. In general, the relic abundance of the scalar field dark matter will depend on the scale of inflation, $H_I$, the equation of state during inflaton oscillations, $w$, the reheating temperature, $\trh$, the mass of the scalar field, $m_\chi$, and its self-interactions, $\lambda_p \chi^p$. Qualitatively,
scalar field fluctuations drive $\langle \chi^2 \rangle$ to large values, which source the field oscillations of $\chi$ after inflation.
Depending on the parameters, $\chi$-oscillations may begin before or after reheating, and their subsequent evolution will depend on their interactions. Once these parameters are specified, it is relatively straightforward to calculate the relic density
of $\chi$. We note that this production is in addition to (and may dominate over) the production of scalar dark matter directly from the inflaton condensate after inflation \cite{ema,Dimopoulos:2006ms,Graham:2015rva,Garny:2015sjg,Tang:2017hvq,Bernal:2018qlk,Ema:2019yrd,Chianese:2020yjo,Ahmed:2020fhc,Kolb:2020fwh,Redi:2020ffc,Ling:2021zlj,MO,Bernal:2021kaj,Barman:2021ugy,Haque:2021mab,cmov,Clery:2022wib,Mambrini:2022uol,Basso:2022tpd,Haque:2023yra,Garcia:2022vwm,Kaneta:2022gug,Kolb:2023dzp,Kaneta:2023kfv,Garcia:2023qab,kkmov}.

Most studies assume that the scalar, $\chi$, is secluded and interacts only through self-interactions and (minimally) through gravity.
As a result, the field is often called a spectator \cite{Markkanen:2018gcw}. The presence or absence of self-interactions
plays a significant role in determining the relic density, as the initial amplitude of $\chi$ oscillations is set by the effective mass of $\chi$. Indeed, in the absence of self-interactions, $\langle \chi^2 \rangle \sim H_I^4/m_\chi^2 \gg H_I^2$ for $m_\chi \ll H_I$. However, the presence of a self-interaction of the form $\lambda_4 \chi^4$ provides a contribution to the effective mass, and we expect $\langle \chi^2 \rangle \sim H_I^2/\sqrt{\lambda_4} \sim H_I^2$ for $\lambda_4 \sim 1$~\cite{Starobinsky:1994bd}, significantly reducing the relic abundance of $\chi$.

On the other hand, the minimal coupling to gravity
ensures that the spectator cannot be entirely
absent from interactions with other fields. The minimal coupling through single graviton exchange provides a coupling between 
the inflaton which (on shell) takes the form $\sigma \phi^2 \chi^2$ with $\sigma  = (m_\phi/2M_P)^2$ \cite{cmov,Clery:2022wib}
for the case of an inflaton potential
$V(\phi)=\frac{1}{2}m_\phi^2\phi^2$, where $M_P = 1/\sqrt{8\pi G} \simeq 2.4 \times 10^{18} \, \rm{GeV}$ is the reduced Planck mass. This term, however,
does not contribute to the effective mass of $\chi$, as it originates from a term in 
the Lagrangian of the form, $\phi^2 (\partial \chi)^2$.\footnote{There is, in fact, a term $\phi^2 \chi^2$ in the 
Lagrangian, but it is highly suppressed with a coupling of order $m_\chi^2/M_P^2$ and will not appreciably affect the generation of the fluctuation in $\chi^2$. } This coupling will also be present if $\chi$ is coupled to curvature ($\mathcal{L} \ni \xi_\chi \chi^2 R$, where $R$ is the Ricci curvature). In this case, a coupling $\xi_\chi (m_\phi^2/M_P^2) \phi^2 \chi^2$ appears when the theory is brought to the Einstein frame \cite{Clery:2022wib}.

Nevertheless, as both the inflaton and spectator are gauge singlets, there is no reason not to expect Planck-suppressed 
operators of the form $\chi^2 \phi^2$, and 
these may greatly affect the generation and subsequent evolution of $\langle \chi^2 \rangle$. In this work, we study the effect of such operators on the relic density of spectator dark matter. 
We consider the simplest case where inflaton oscillations are driven by a 
quadratic term about minimum of the inflaton potential ($w=0$), motivated by a simple inflationary model \cite{Staro}. Once the inflaton potential parameters are fixed from CMB 
observations, we have three independent free parameters: 
the spectator mass, $m_\chi$, the spectator coupling to the inflaton, $\sigma$, and the 
reheating temperature, $\trh$. 
The relative values of these parameters will determine when the effective spectator 
mass is dominated by its bare mass, when the spectator oscillations begin, and when reheating is achieved. The corresponding values of the cosmological scale factor for these events are denoted respectively
by $a_\chi$, $a_{\rm osc}$, and $a_{\rm RH}$.  Without either self-interactions or couplings to other fields, as already shown in \cite{Cosme:2020nac,Lebedev:2022cic}, there can be a severe overproduction of scalar dark matter, as we will also see below.  Adding a coupling to the inflaton greatly relaxes this constraint and opens up the allowed values of $m_\chi$ and $\trh$.

We will show that the parameter space begins to open for Planck-suppressed couplings\footnote{In the theory with a non-minimal coupling to curvature, we can make the associations ${\tilde \sigma} = - \xi_\chi$ in the effective mass \cite{Clery:2022wib}. This association is only valid when $\xi_\chi \chi^2 \ll M_P^2$.} $\sigma = {\tilde \sigma} m_\phi^2/M_P^2$, with ${\tilde \sigma} \gtrsim 10^{-7}$. Furthermore, for ${\tilde \sigma} \gtrsim \frac{9}{16}$, particle production becomes dominated by inflaton scattering through either single graviton exchange or the contact interaction between $\chi$ and $\phi$. In this case the allowed range on the mass of $\chi$ 
is very large, $\sim 0.01 - \sim 10^{13}$~GeV and the allowed range on $\trh$ also opens, reaching roughly $0.004 - 10^{15}$~GeV. This can be accomplished with couplings ${\tilde \sigma}$ between 0.004 and 50, so that constraints from isocurvature perturbations are avoided and non-perturbative production from scattering can be neglected. 

In what follows, we first define the model of inflation, the coupling to the scalar, and the cosmological evolution of the system in Section \ref{sec:model}.  In Section \ref{sec:range}, we define our model parameters and consider the production of long wavelength modes (important for relatively small couplings), combined with the gravitational production of $\chi$ through scattering of inflaton field. We also discuss the constraints from isocurvature perturbations. Our results are collected in Section \ref{sec:scenarios}, where we distinguish between large and small coupling scenarios. A discussion and summary of our results is given in Section \ref{sec:discussion}.

\section{Model and Field Evolution after Inflation} 
\label{sec:model}
We consider a model with two real scalar fields: the inflaton, $\phi$, and the spectator, $\chi$.  The action is given by
\begin{equation}
\begin{aligned}
    \label{eq:action}
    \mathcal{S}\; = \; \int d^4x \sqrt{-g} \bigg[ \frac{1}{2} \partial^{\mu} \phi \partial_{\mu} \phi & - V(\phi) + \frac{1}{2} \partial_{\mu} \chi \partial^{\mu} \chi \\
    & - \frac{1}{2} m_\chi^2 \chi^2 - \frac{1}{2}\sigma \phi^2 \chi^2 \bigg] \, ,
\end{aligned}
\end{equation}
where $V(\phi)$ is the inflationary potential, $m_{\chi}$ is the bare mass of the spectator field, and $\sigma$ is a direct coupling between the inflaton and spectator field. Although this coupling appears renormalizable, we assume it is an effective and Planck-suppressed coupling. Therefore, we expect $\sigma$ to be of order $m_\phi^2/M_P^2$ in analogy with the on-shell coupling from single graviton exchange \cite{cmov,Clery:2022wib}.  We use the metric signature $(+, -, -, -)$. 

Varying this action with respect to $\phi$ and $\chi$, the equations of motion are\footnote{We neglected the $\sigma \langle \chi^2 \rangle \phi$ contribution because it is of order $\sigma \langle \chi^2\rangle \ll m_{\phi}^2$.
} 
\begin{equation}
    \label{eq:eomphi}
    \ddot{\phi} + 3H \dot{\phi} + \frac{dV(\phi)}{d\phi} \; \simeq \; 0 \, ,
\end{equation}
\begin{equation}
    \label{eq:eomS}
    \ddot{\chi} + 3H \dot{\chi} + m_{\chi, \rm eff}^2 \chi \; = \; 0 \, ,
\end{equation}
where $H = \frac{\dot{a}}{a}$ is the Hubble parameter, and the effective mass of the spectator field $\chi$ is given by $m_{\chi, \rm eff}^2 \; = \; m_\chi^2  + \sigma \langle \phi^2 \rangle$. When accelerated expansion ends, the inflaton begins a series of oscillations, dominating the energy density of the Universe until reheating, when the radiation density (produced, for example, by inflaton decays or scatterings
\cite{gkmo1,gkmo2}) equals the energy density in the inflaton condensate. Similarly, the spectator field, $\chi$, will also begin its series of oscillations when $m t \simeq 1$ or 
$m_{\chi, \rm eff}(a) = \frac32 H(a)$ in a matter-dominated (inflaton-dominated) universe, occurring at $a=a_{\rm osc}$. The redshifting of the energy density in $\chi$ is highly dependent on the 
content of the Universe, and we show how the relative values of $a_{\rm osc}$ and $a_{\rm RH}$ are crucial in determining the relic density of $\chi$ at the present time.

For numerical analysis, we consider the Starobinsky model of inflation \cite{Staro},
\beq
V(\phi) \; = \; \frac34 m_{\phi}^2 M_P^2 \left(1 - e^{-\sqrt{\frac{2}{3}} \phi }
 \right)^2 \, ,
 \label{starpot}
\eeq
which, when expanded about the minimum, behaves as
\begin{equation}
    \label{eq:infpotmin}
    V(\phi) \; \simeq \; \frac12 m_\phi^2 \phi^2 \, , \qquad \phi \lesssim M_P \, ,
\end{equation}
with $m_\phi = 2 H_I$, where $H_I$ is the Hubble parameter during inflation. The normalization of the inflaton potential can be obtained from the overall amplitude of scalar perturbations,
$A_{S^*} = 2.1 \times 10^{-9}$ \cite{Planck:2018jri} and
\begin{equation}
   \label{eq:lambda}
   \frac{m_\phi^2}{M_P^2} \; \simeq \; \frac{24 \pi^2 A_{S^*}}{ N_*^2} \, .
\end{equation}
For a nominal choice of $55$ $e$-folds,\footnote{We take $55$ $e$-folds for simplicity. See~\cite{egnov} for a detailed analysis of the Starobinsky model that constrains the number of $e$-folds while accounting for reheating.} for the Starobinsky potential, one finds $m_\phi \simeq 1.25 \times 10^{-5} M_{P} \simeq 3 \times 10^{13}~{\rm GeV}$ \cite{eno6,EGNO5,egnov,Ema:2024sit}. Additionally, the scalar tilt, $n_s = 0.965$, and scalar-to-tensor anisotropy ratio, $r = 0.0035$, are in excellent agreement with Planck results \cite{Planck:2018jri}.

The solution to Eq.~(\ref{eq:eomphi}) is well known and leads to harmonic oscillations of $\phi$. The energy density can be expressed in terms of the amplitude of the oscillations and $\rho_\phi = \frac12\dot \phi^2+V(\phi)=\rho_{\rm end} (a_{\rm end}/a)^{3}$. Here, the scale factor at the end of inflation is denoted as $a_{\rm end}$, which is derived from the fact that at $\aend$, $V(\phi_{\rm end})=\dot \phi_{\rm end}^2$ and $\rho_{\rm end} = \frac32 V(\phi_{\rm end})$.\footnote{Equivalently, inflation ends when the accelerated expansion stops and $\ddot{a}(t_{\rm end}) = 0$.} The inflaton field value at the end of inflation is denoted as $\phi_{\rm end} = \phi(a_{\rm end})$, and similarly, $H_{\rm end} = H(a_{\rm end})$.

Supposing that $\rho_\phi$ dominates the energy budget of the Universe until reheating is complete, the Hubble parameter scales as $H(a) \propto \sqrt{\rho_\phi} \sim a^{-\frac{3}{2}} $ for $a_{\rm end} < a < a_{\rm RH}$.
It is then convenient to rewrite the equation of motion Eq.~(\ref{eq:eomS}) for $\chi$ as
\begin{equation}
   \chi'' + \frac52  \frac{\chi'}{a} + 4 \tilde{\sigma} \frac{\chi}{a^{2}} \; = \; 0 \, , \qquad \tilde{\sigma} = \sigma \frac{ M_P^2}{m_\phi^2} \, ,
   \label{Eq:chidiff}
\end{equation}
where $' \equiv d/da = (aH)^{-1}d/dt$,  and we assumed that the contribution from the bare mass term $m_{\chi}^2$ is negligible. We also used the relation, $4H_{\rm end}^2 M_P^2 = m_{\phi}^2 \phi_{\rm end}^2$ at the end of inflation.

If we choose a general ansatz of the form $\chi(a) \propto a^{-A+i B}$, the solution for real $B$ is given by: 
\begin{equation}
\begin{aligned}
&&
\chi(a)=
\chi_{\rm end}\left(\frac{\aend}{a}\right)^\frac34
\left[
\cos\left(B\log\left[\frac{a}{\aend}\right]\right)
\right.
\\
&&
+\left.
\frac{3}{4B}\sin\left(B\log\left[\frac{a}{\aend}\right]\right)
\right]
\,,
\label{eq:chi(a)}
\end{aligned}
\end{equation}
with  $\chi_{\rm end} = \chi(\aend)$, $A = \frac{3}{4}$, 
and $B = 2 \sqrt{ {\tilde \sigma}-\frac{9}{64}}$. 
The constants of integration have been set by assuming $\chi({\aend}) = \chi_{\rm end}$ and $\chi^\prime(\aend) = 0$.  Real $B$ then requires ${\tilde \sigma} \ge 9/64$. For smaller ${\tilde \sigma}$, the solution is no longer oscillatory and takes the form
\beq
\chi(a) = \chi_{\rm end} \frac{1}{2c} \left(\frac{\aend}{a}\right)^\frac{c+3}{4} \left( c-3 + \left(\frac{a}{\aend }\right)^\frac{c}{2} (c+3) \right) \, ,
\eeq
with $c= 8 \sqrt{\frac{9}{64} - {\tilde \sigma}}$. These solutions break down when the bare mass term for $\chi$ cannot be neglected in Eq.~(\ref{Eq:chidiff}). This occurs when ${\tilde \sigma}$ is very small, or at large $a$, when $\phi^2$ is small.  Then, Eq.~(\ref{Eq:chidiff}) is modified (by replacing the third term with $4 {\tilde \sigma}/a^2 \to m_\chi^2 a/H_{\rm end}^2 \aend^3$) and the spectator field scales like a matter field, with $\rho_\chi \propto a^{-3}$ and the amplitude of $\chi \propto a^{-3/2}$. We also discuss this possibility below.

\section{Production Regimes} 
\label{sec:range}
In this work, we focus on a broad range of $\tilde{\sigma}$ up to $\tilde{\sigma} \lesssim 50$, where the broad parametric resonance effects become significant and a full non-perturbative analysis is required~\cite{Garcia:2021iag,Garcia:2022vwm}. However, as shown in \cite{Garcia:2021iag}, one can treat the dark matter production perturbatively up to this value (including the narrow resonance regime with $\tilde{\sigma} \sim \mathcal{O}(1)$). Most importantly, we demonstrate that a direct coupling between the inflaton and dark matter field opens up a broad parameter space, allowing for a wide range of dark matter masses and reheating temperatures, while successfully avoiding isocurvature constraints.\footnote{We note that similar effects can be achieved by introducing a non-minimal coupling between the dark matter field and gravity~\cite{Garcia:2023qab}.}

\subsection{Long Wavelength Contribution}
When $\tilde{\sigma} \lesssim \mathcal{O}(1)$, the production of the spectator field, $\chi$, is primarily dominated by the long wavelength contribution (superhorizon modes). We assume that during inflation, the energy density of the spectator field is largely subdominant, and inflation is driven solely by the inflaton, $\phi$. At the end of inflation, we further assume that the expectation value of the spectator field, $\chi$, has reached its asymptotic value given by \cite{fluc}
\begin{equation}
    \label{eq:exptvalS}
    \langle \chi^2 \rangle_{\rm end} \; \simeq \frac{3H_{\rm end}^4}{8 \pi^2 m^2_{\chi, \rm eff}(a_{\rm end})} \, .
\end{equation}
This result corresponds to the spectator field variance that is averaged over long wavelength superhorizon modes, and we show the derivation of this result using the stochastic approach in Appendix~\ref{app:fp}. 
 
Assuming that in a horizon patch, the field is approximately homogeneous, the field value after inflation is given by $\chi_{\rm end} = \sqrt{ \langle \chi^2 \rangle_{\rm end}}$. When $\sigma \phi^2(a) \gg m_\chi^2$, one can show that the energy density of the spectator field evolves as
\beq
 \rho_\chi(a) \; = \;  \frac{1}{2} m^2_{\chi, \rm eff} \chi^2(a) \; \simeq \;  \frac{3H_{\rm end}^4}{16\pi^2}  \left(\frac{\phi(a)}{\phi_{\rm end}} \right)^2 \left(\frac{a_{\rm end}}{a}\right)^{\frac32} \, .
\eeq
We note that because the inflaton field scales as $\phi(a)^{2} \propto a^{-3}$ and $\chi \propto a^{-3/2}$, this leads to an overall spectator field energy density scaling of $\rho_\chi(a) \propto a^{-\frac92}$. 
This means that $\rho_\chi$ redshifts faster than $\rho_\phi$,
and will not dominate the energy budget of the Universe before reheating is complete.

We note that for large values of the effective mass, the fluctuations are cut off exponentially with $\langle \chi^2 \rangle \propto e^{-c \pi (m^2_{\chi, \rm eff}/H_{\rm end}^2-\frac94)}$, where $c \sim \mathcal{O}(1)$ model-dependent constant~\cite{Chung:1998bt, Markkanen:2016aes, Chung:2018ayg, Racco:2024aac}.\footnote{During inflation, the asymptotic value of $\langle \chi^2 \rangle$ is determined by $H_I$, but numerical studies indicate that the subsequent evolution and density of $\chi$ is better fit using $H_{\rm end}$ \cite{paper2}.} Therefore, the exponential suppression becomes significant when 
\begin{equation}
    \frac{m^2_{\chi, \rm eff}(a_{\rm end})}{H_{\rm end}^2} \gtrsim \frac{9}{4}\, ,~~m^2_{\chi, \rm eff}(a_{\rm end}) \; = \; m^2_{\chi} + 4 \tilde{\sigma} H^2_{\rm end} \, , 
\end{equation}
where we used $\phi^2_{\rm end} = 4 H_{\rm end}^2 M_P^2/m_{\phi}^2$. This implies that the exponential suppression needs to be accounted for when $m_{\chi}^2 \gtrsim \frac{9}{4}H_{\rm end}^2$ when $4 \tilde{\sigma} H_{\rm end}^2 \ll m_{\chi}^2$, and $\tilde{\sigma} \gtrsim \frac{9}{16}$ when $4 \tilde{\sigma} H_{\rm end}^2 \gg m_{\chi}^2$. Due to the exponential suppression, the long wavelength contribution becomes subdominant for large values of $\tilde{\sigma} \gg \frac{9}{16}$, and the contribution from the gravitational scattering begins to dominate. 

\subsection{Gravitational Production After Inflation}
When the direct coupling between the inflaton and spectator field becomes large, i.e., $\tilde{\sigma} \gtrsim \frac{9}{16}$, the universal gravitational interactions between the inflationary and dark sectors start dominating the dark matter production. To compute this gravitational production of dark matter during reheating, we expand the spacetime metric around the flat Minkowski background, with $g_{\mu \nu} \simeq \eta_{\mu \nu} + \frac{2}{M_P}h_{\mu \nu}$, where $h_{\mu \nu}$ is the canonically-normalized linear perturbation. Combining gravitational interactions with the action in Eq.~(\ref{eq:action}), we find the following interaction terms (see e.g. \cite{hol})
\begin{equation}
    \mathcal{L}_{I} \; = \; -\frac{1}{M_P} h_{\mu \nu} \left(T_{\phi}^{\mu \nu} + T_{\chi}^{\mu \nu} \right) - \frac{1}{2} \sigma \phi^2 \chi^2 \, ,
    \label{eq:interaction}
\end{equation}
where $T_{\phi}^{\mu \nu}$ and $T_{\chi}^{\mu \nu}$ are the energy-momentum tensors of the inflaton and the spectator dark matter field, respectively, given by
\begin{equation}
    T_{S = \phi, \, \chi}^{\mu \nu} \; = \; \partial^{\mu} S \partial^{\nu} S - g^{\mu \nu} \left[\frac{1}{2} \partial^{\alpha}S \partial_{\alpha} S - \frac{1}{2} m_{\rm{eff}}^2 S^2 \right]  \, ,
\end{equation}
where $m_{\rm eff}$ is the effective mass. The field perturbation $h_{\mu \nu}$ characterizes a massless spin-2 (canonically-normalized) graviton. Its propagator in de Donder gauge that carries the momentum $k$
can be expressed as~\cite{Giudice:1998ck}
\begin{equation}
    \label{app:gravprop}
    \Pi^{\mu \nu \rho \sigma }(k)\,=\, \dfrac{\eta^{\rho \nu} \eta^{\sigma \mu } + \eta^{\rho \mu} \eta^{\sigma \nu } - \eta^{\rho \sigma} \eta^{\mu \nu }   }{2k^2} \, .
\end{equation}
The matrix element that corresponds to the final state of a pair of dark matter particles, can be computed by treating the inflaton condensate as a collection of particles, leading to dark matter production via the $s$-channel scattering process~\cite{MO,cmov,Clery:2022wib}.

Importantly, because the effective mass of dark matter is $\tilde{\sigma}$-dependent, it will affect the scattering process and lead to kinematic blocking when $m_{\chi, \rm eff}^2 \gtrsim m_{\phi}^2$, delaying the scattering processes. However, even though the channel $\phi \phi \rightarrow h_{\mu \nu} \rightarrow \chi \chi$ is sensitive to the reheating temperature $T_{\rm RH}$ at the early stages of reheating~\cite{cmov}, as the value of a direct coupling $\tilde{\sigma}$ increases, the scattering contribution arising from the $\sigma \phi^2 \chi^2$ coupling begins to dominate dark matter production, leading to efficient particle production. We discuss this in detail in the next section.

\subsection{Model Parameters}
We summarize the key parameters for the Starobinsky model of inflation with $N_* = 55$ $e$-folds that we use for our numerical
results, along with the field and energy density dependence on the scale factor $a$: 
\begin{equation}
\begin{aligned}
    m_\phi = 1.25 \times 10^{-5} M_P \, ,\qquad  & \phi_{\rm end} = 0.61\,M_P \, ,  \\
  H_{\rm end} = 3.1 \times 10^{-6} M_P \, ,  \qquad & |\dot{\phi}_{\rm end}| = \sqrt{2} H_{\rm end} \,M_P \, ,  \\
    \chi \propto a^{-\frac34} \, ,\qquad & \phi \propto a^{-\frac32} \, ,  \\
    \rho_\chi \propto a^{-\frac92} \, , \qquad & \rho_{\phi} \propto a^{-3} \, .
    \label{summ}
\end{aligned}
\end{equation}
The scaling for $\chi$ applies when $a_{\rm osc} < a < a_\chi$, that is after oscillations begin and before the effective mass is dominated by the bare mass. Subsequently, $\rho_\chi \propto a^{-3}$ as its mass is constant. 

As noted earlier, there are three important scale parameters: $a_{\rm osc}$, $a_\chi$, and $a_{\rm RH}$. Their value relative to $a_{\rm end}$ can be determined from their definitions: $m_{\chi, \rm eff}(a_{\rm osc}) = \frac32 H(a_{\rm osc})$, $\sigma \phi^2(a_\chi) = m_\chi^2 $, and $\rho_\phi(a_{\rm RH}) = \rho_{\rm end}(a_{\rm end}/a_{\rm RH})^3 =  \rho_{\rm radiation} = \alpha \trh^4$, where $\alpha = g_{\rm RH} \pi^2/30$ and $g_{\rm RH} = 427/4$ is the number of Standard Model degrees of freedom.\footnote{This assumes $\trh > m_t$ (the top quark mass). For lower $\trh$, $g_{\rm RH}$ must be appropriately reduced.}
The solutions are given by
\begin{align}
    \left(\frac{a_{\rm osc}}{a_{\rm end}}\right)^3 & =
{\rm Max}\left(\frac{9 H_{\rm end}^2}{4 m_\chi^2} \left(1 - \frac{16}{9}{\tilde \sigma} \right),1\right) \label{aoae} \, ,\\
     \left(\frac{a_{\chi}}{a_{\rm end}}\right)^3 & =  {\rm Max}\left(\frac{\sigma \phi_{\rm end}^2}{m_\chi^2 },1\right) = {\rm Max}\left(\frac{4 {\tilde \sigma} H_{\rm end}^2}{ m_\chi^2},1\right) \label{acae} \, ,\\
      \left(\frac{a_{\rm RH}}{a_{\rm end}}\right)^3 & =  \frac{3 H_{\rm end}^2 M_P^2}{\alpha \trh^4 } \label{arhae} \, .
\end{align}
Note that for $\tilde \sigma \gtrsim \frac{9}{16}$, the oscillations begin at the end of inflation, $\aosc=\aend$. For smaller $\tilde \sigma$, 
$(\aosc/\aend)^3 \simeq (1.1\times 10^{13}~{\rm GeV}/m_\chi)^2$. Similarly, 
$(\achi/\aend)^3
\simeq {\tilde \sigma}
(1.5\times 10^{13}~{\rm GeV}/m_\chi)^2 $.

The order of these scale factors depends on $\sigma$, $m_\chi$, and $\trh$,
assuming that $\phi_{\rm end}$ and $H_{\rm end}$ are set by the inflationary model.  Thus, we have
\begin{align}
 & \qquad{\tilde  \sigma}  < \frac{9}{32} \, , &a_\chi < a_{\rm osc} \, , \label{achiltaosc} \\
    & \qquad {\tilde  \sigma} < \frac34 \frac{m_\chi^2 M_P^2}{\alpha \trh^4} \, , &a_\chi < a_{\rm RH} \, , \label{acltarh}\\
   & \qquad {\tilde  \sigma} < \frac{9}{16}  \left( 1- \frac{4 M_P^2 m_\chi^2}{3 \alpha \trh^4}\right)\, , &a_{\rm RH} < a_{\rm osc} \, .
   \label{arhltaosc}
\end{align}
Each of these cases implies a different evolution for
$\rho_\chi$. We will discuss them one by one in the following.

\subsection{Isocurvature Constraints}
We also discuss the isocurvature constraints. The current isocurvature power spectrum constraints from Planck are given by
\begin{equation}
    \beta_{\rm iso} \; \equiv \; \frac{\mathcal{P}_{\mathcal{S}}(k_*)}{\mathcal{P}_{\mathcal{R}}(k_*) + \mathcal{P}_{\mathcal{S}}(k_*)} < 0.038 \, ,
\end{equation}
at $95 \%$ CL with the pivot scale $k_* = 0.05 \, \rm{Mpc}^{-1}$~\cite{Planck:2018jri}, where $\mathcal{P}_{\mathcal{R}}$ represents the curvature spectra and $\mathcal{P}_{\mathcal{S}}$ denotes the isocurvature spectra. During inflation, if dark matter is light compared to the inflationary scale, it leads to the production of isocurvature modes in tension with the current Planck constraints~\cite{Chung:2004nh, Garcia:2023awt}. 

The curvature spectrum of a massive scalar field can be expressed as
\begin{equation}
   \mathcal{P}_{\mathcal{R}}(k) \; = \; \frac{H_I^2}{4\pi^2} \left(\frac{k}{a H_I} \right)^{3-2\sqrt{\frac{9}{4} - \frac{m^2_{\chi, \rm eff}}{H_I^2}}} \, ,
\end{equation}
when $m^2_{\chi, \rm eff}/H_I^2 \lesssim 9/4$, and the isocurvature power spectrum is given by
\begin{equation}
\mathcal{P}_{\mathcal{S}}(k) \; = \; \frac{k^3}{2 \pi^2 \langle \rho_\chi \rangle^2} \int \mathrm{d}^3 x\left\langle\delta \rho_\chi(x) \delta \rho_\chi(0)\right\rangle e^{-i \boldsymbol{k} \cdot \boldsymbol{x}},
\end{equation}
where $\rho_{\chi}$ is the dark matter density and $\delta \rho_{\chi}$ is the dark matter fluctuation. Because the dark matter energy density scales as $\rho_{\chi}(x) \propto \chi^2(x)$, and assuming Gaussianity, we can use the approximation
\begin{equation}
\frac{\langle\delta \rho_{\chi}(x) \delta \rho_{\chi}(0)\rangle}{\langle\rho_{\chi}\rangle^2}=\frac{\left\langle\chi^2(x) \chi^2(0)\right\rangle-\left\langle\chi^2\right\rangle^2}{\left\langle\chi^2\right\rangle^2}=2 \frac{\langle\chi(x) \chi(0)\rangle^2}{\left\langle\chi^2\right\rangle^2} \, .
\end{equation}
This allows us to recast the isocurvature power spectrum in the following form
\begin{equation}
\begin{aligned}
   \mathcal{P}_{\mathcal{S}}(k) & =  \frac{k^3}{2 \pi^2} \frac{2}{\left\langle\chi^2\right\rangle^2} \int d^3 x e^{-i \boldsymbol{k} \cdot \boldsymbol{x}} \langle\chi(x) \chi(0)\rangle^2 \\
   &=\frac{k^3}{2 \pi^2} \frac{2}{\left\langle\chi^2\right\rangle^2} \int \frac{d^3 q}{(2 \pi)^3} P(q) P(k-q) \, .
\end{aligned}   
\end{equation}
If we directly compute this integral, we obtain~\cite{Redi:2022zkt}
\begin{equation}
    \mathcal{P}_{\mathcal{S}}(k) \; \simeq \; \frac{8 m_{\chi, \rm eff}^2}{3H_{\rm end}^2} \left(\frac{k}{a_{\rm end} H_{I}} \right)^{\frac{4 m_{\chi, \rm eff}^2}{3H_{I}^2}} \, .
\end{equation}
Using this expression, we observe that the modes that exit the horizon early during inflation will be strongly suppressed. Therefore, for the $k_*$ mode that exits after $N_* = 55$ $e$-folds of inflation, we can use the approximation
\begin{equation}
    \mathcal{P}_{\mathcal{S}}(k_*) \; \simeq \; \frac{8 m_{\chi, \rm eff}^2}{3H_I} \exp\left(\frac{-4N_* m^2_{\chi, \rm eff}}{3H_I^2} \right) \, .
\end{equation}
If we now use the $N_* = 55$ $e$-folds, we find the constraint
\begin{equation}
    m_{\chi, \rm eff} (t_*) \gtrsim 0.5 H_{I} \simeq H_{\rm end} \, ,
\end{equation}
where we used the fact that for the Starobinsky model of inflation, $H_I \simeq 2H_{\rm end}$. This constraint implies that 
\begin{align}
    m_{\chi} & \gtrsim H_{\rm end} \, , \qquad \sigma \phi_{*}^2 \ll m^2_{\chi, \rm eff} \, , \\
    \tilde{\sigma} & \gtrsim 0.003 \, , \qquad \sigma \phi_{*}^2 \gg m^2_{\chi, \rm eff} \, ,
\end{align}
where in the second line, we used the ratio $\phi_*^2/\phi_{\rm end}^2 \simeq 75$ for the Starobinsky model of inflation.

A fully numerical computation of the isocurvature constraint for the Starobinsky model of inflation, that was performed in Ref.~\cite{Garcia:2023awt}, shows that the isocurvature constraints are given by
\begin{align}
    m_{\chi} & \gtrsim 1.1 H_{\rm end} \, , &\sigma \phi_{*}^2 \ll m^2_{\chi, \rm eff} \, , \\
    \tilde{\sigma} & \gtrsim 0.004 \, , &\sigma \phi_{*}^2 \gg m^2_{\chi, \rm eff} \, ,
    \label{eq:sigmatildeiso}
\end{align}
which are in excellent agreement with the analytical results. We use the fully numerical results when discussing the dark matter constraints. However, the analytical procedure carried out in this section is general and can be easily applied to various models of inflation. 

\section{Scenarios}
\label{sec:scenarios}
\subsection{Case I: $\sigma \phi_{\rm end}^2 \ll m_{\chi}^2$}

When $\sigma$ is very small (or absent), the bare mass term comes to dominate early (and will then always dominate at later times), and $a_\chi < a_{\rm osc}, a_{\rm RH}$. The spectator field starts oscillating when $\frac32 H(a_{\rm osc}) \simeq m_{\chi}$. Because the Hubble parameter scales as $H(a) \propto a^{-\frac{3}{2}}$, the oscillations start before reheating if $\frac34 \frac{\rho_{\rm RH}}{M_P^2}< m_\chi^2$, or $3 \alpha \trh^4 < 4 m_\chi^2 M_P^2$, when $a_{\rm osc}/\aend = (3 H_{\rm end}/2 m_\chi)^{2/3}$. We refer to this case as IA, where $a_{\rm \chi} < a_{\rm osc}  < a_{\rm RH}$. 

The initial value when the field starts oscillating is given by $\chi(a_{\rm osc}) = \sqrt{\langle \chi^2 \rangle_{\rm end}}$. We want to compute the energy density of the spectator field at the time of reheating.
By combining Eqs.~(\ref{eq:exptvalS}), (\ref{aoae}) and (\ref{arhae}), we find
\begin{equation}
\begin{aligned}
   &\rho_{\chi}(a_{\rm RH}) \; =  \; \frac{1}{2}m_\chi^2\langle \chi^2\rangle_{\rm end}
\left(\frac{\aosc}{\aend}\right)^3\left(\frac{\aend}{\arh}\right)^3  \\
& \; = \;  \frac{9}{64\pi^2} H_{\rm end}^4 \frac{\alpha \trh^4}{M_P^2({m_\chi^2 + \sigma \phi_{\rm end}^2}) } \left(1 - \frac{16}{9} \tilde \sigma \right)  \\
 & \; = \; \frac{9 \alpha H_{\rm end}^4\trh^4}{64 \pi^2m_\chi^2M_P^2} \, , \qquad \sigma \phi_{\rm end} \ll m_\chi^2; \, {\tilde \sigma \ll 1}\, .
  \label{rhochiarh}
\end{aligned}
\end{equation}

\begin{figure}[t!]
  \centering
\includegraphics[width=0.48\textwidth]{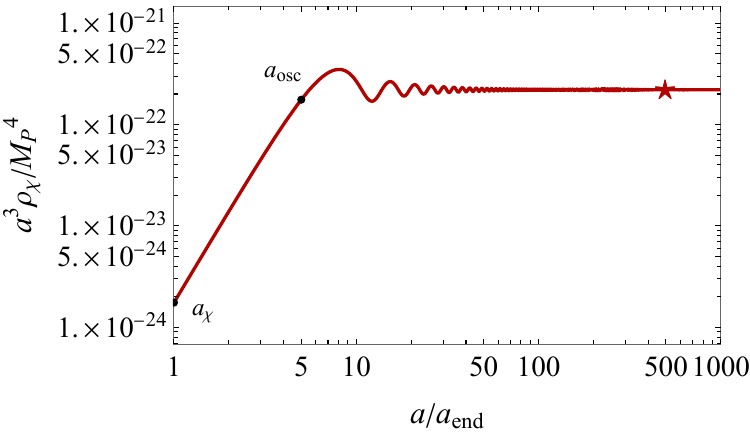}
  \caption{\em \small {
Energy density evolution of the spectator field $\rho_\chi a^3$.
Here we assumed $\sigma \phi_{\rm end}^2 \ll m_\chi^2$, $m_\chi = 10^{12}$~GeV, and $\trh = 77$~GeV. The star shows the relic density obtained from Eq.~(\ref{rhochiarh})} 
}
  \label{fig:enden1}
\end{figure}

The numerical evolution of $\rho_\chi$ is shown in Fig.~\ref{fig:enden1} when multiplied by $a^3$. The choice of parameters is given in the caption and have been chosen to yield a present density of $\Omega_\chi h^2 = 0.12$. The dots indicate the values of $a_\chi = 1$, $a_{\rm osc} = 5$. Since the reheating temperature in this case is low, $a_{\rm RH} = \mathcal{O}(10^{18})$, far off the scale of the plot (values of $a$ are all relative to $\aend$). The star shows the result from Eq.~(\ref{rhochiarh}) and is in excellent agreement with the numerical result. 

To obtain the present-day relic abundance of dark matter, the energy density in Eq.~(\ref{rhochiarh}) must be redshifted to the present
by 
\beq
\left(\frac{a_{\rm RH}}{a_0}\right)^3 = \left( \frac{T_0}{\trh}\right)^3 \frac{g_0}{g_{\rm RH}} \, ,
\label{today}
\eeq
where $T_0$ is the present temperature of the CMB, and 
$g_0 = (43/4)(4/11)$.
The fraction of critical density at the present time is then
given by \cite{mybook}
\begin{equation}
    \begin{aligned}
        &&
    \frac{\Omega_{\chi} h^2}{0.12} \; = \; \frac{\rho_{\chi}(a_0)}{\rho_c }~\frac{h^2}{0.12} \; \simeq 10^{47} \frac{\rho_\chi(a_0)}{\rm GeV^4}\\
    &&
    \simeq \; 4.9 \times 10^7~{\rm GeV}^{-1}~ \frac{\rho_{\chi}(a_{\rm RH}) }{\trh^3}  \, ,
     \label{eq:dmdensity}
    \end{aligned}
\end{equation}
where we used
\beq
\rho_c = 8.1\times 10^{-47}\,h^2 \,{\rm GeV^4} = 1.054 \times 10^{-5} \, h^2 \, \rm{GeV} \, \rm{cm}^{-3}\,.
\eeq
Combining these results, we find
\begin{align}
    &\frac{\Omega_\chi h^2}{0.12} \; \simeq  \; 2.5 \times 10^7 \, \mathrm{GeV}^{-1} \frac{H_{\rm end}^4 T_{\rm RH}}{M_P^2 ({m_\chi^2 + \sigma \phi_{\rm end}^2}) }\left(1 - \frac{16}{9} \tilde \sigma\right) \nonumber \\
   & \; \simeq  \; \frac{\trh}{77~{\rm GeV}}\left(\frac{10^{12}~{\rm GeV}}{m_\chi}\right)^2 \, ,  \qquad \sigma \phi_{\rm end} \ll m_\chi^2; \, {\tilde \sigma \ll 1}\, . 
    \label{Eq:omegaIA}
\end{align}

\begin{figure*}[ht!]
  \centering
  \hspace*{-5mm}
\subfigure{\includegraphics[scale=0.34]{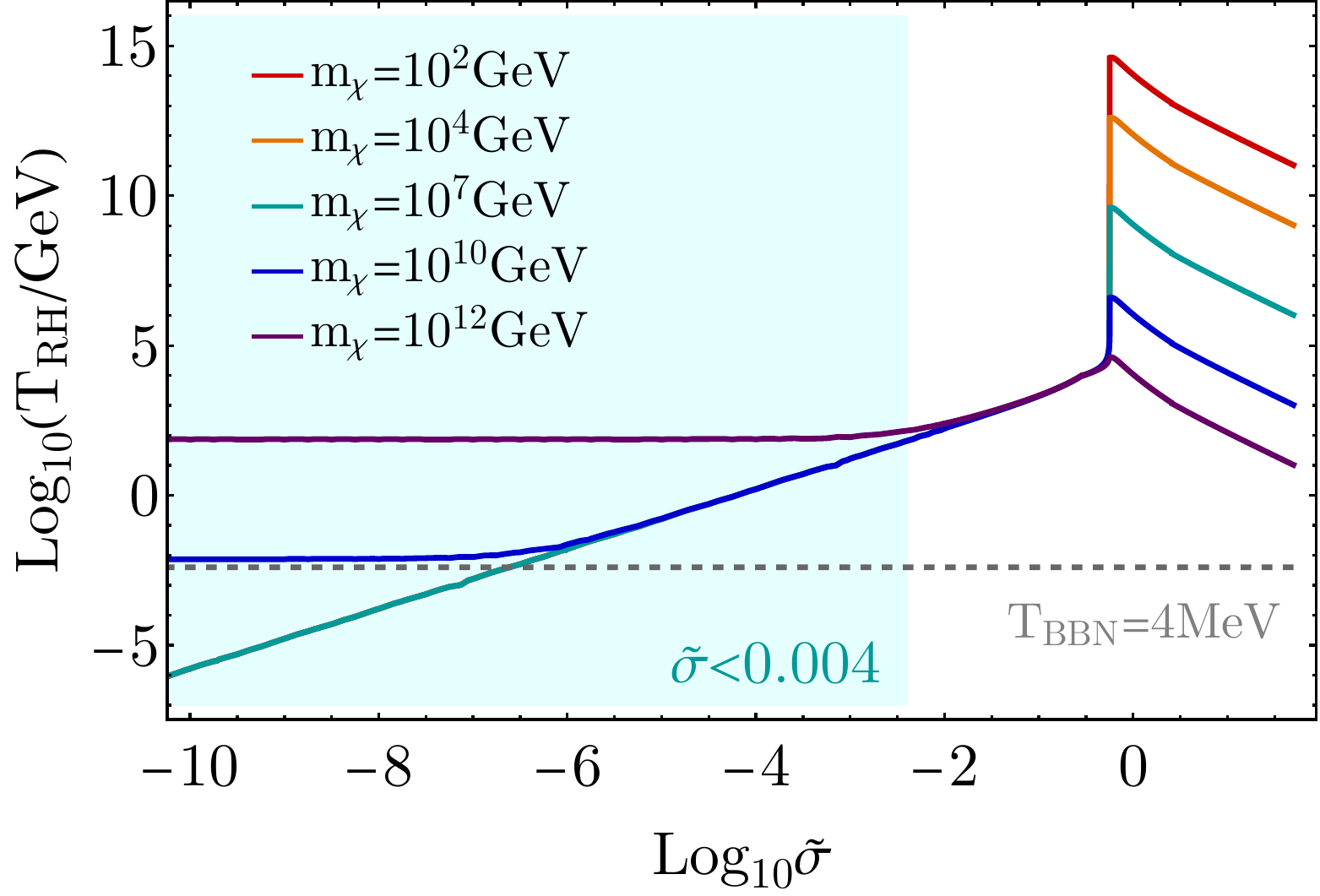}}\quad
\subfigure{\includegraphics[scale=0.335]{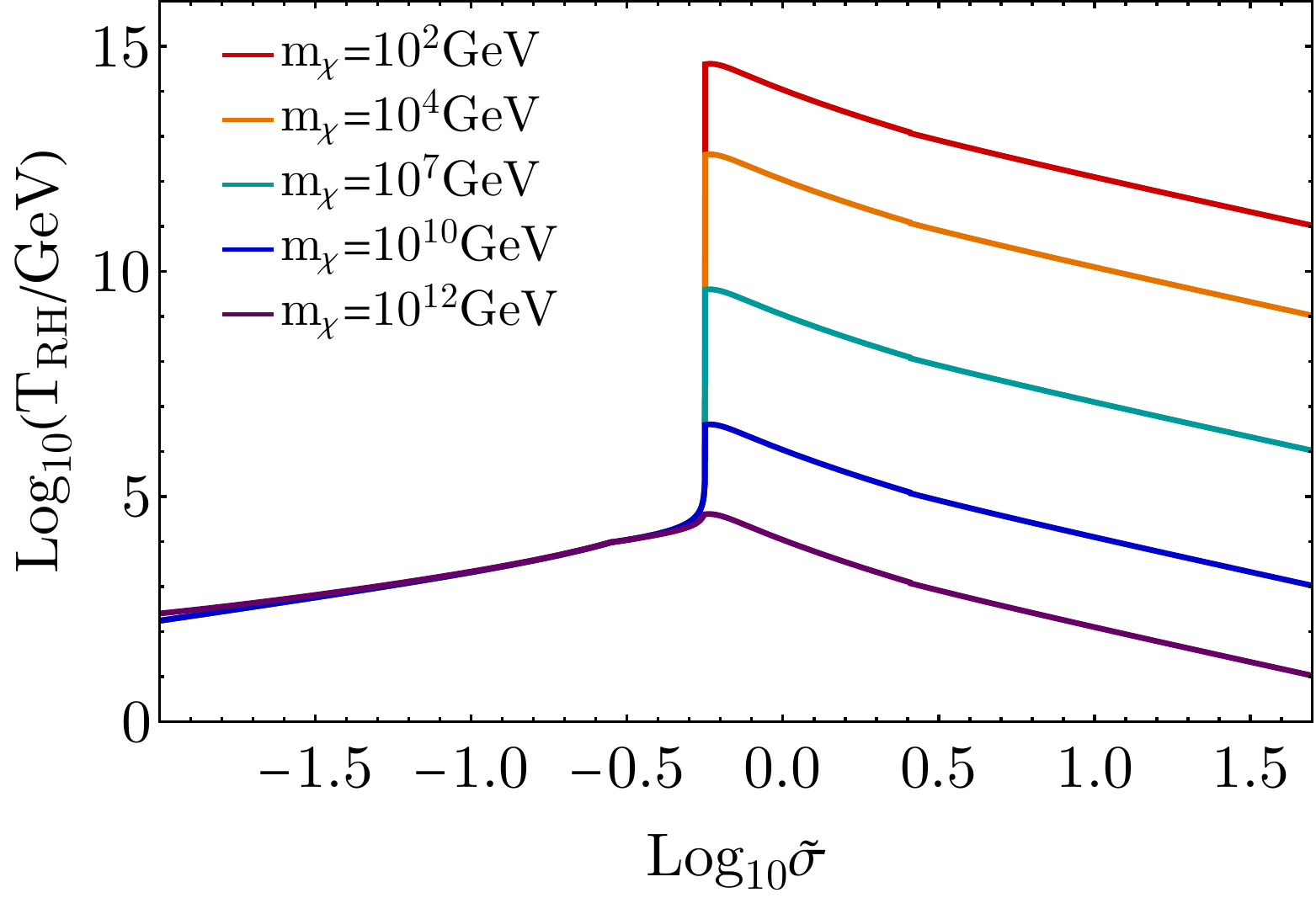}}
  \caption{\em \small {$T_{\rm RH}$ vs $\tilde{\sigma}$  satisfying $\Omega_{\chi}h^{2}=0.12$ for $m_\chi = 10^{2}$(red), $10^{4}$ (orange), $10^{7}$ (green), $10^{10}$ (blue), $10^{12}$ (purple) GeV. The gray dashed line on the left panel shows the BBN bound on $T_{\rm RH}$. The cyan shaded region in the left panel is excluded by the isocurvature constraint in Eq.~(\ref{eq:sigmatildeiso}). The right panel magnifies the range $10^{-2}\leq \tilde{\sigma} \leq50$ of the left panel.
}
} 
  \vspace*{-1.5mm}
\label{fig:test1}
\end{figure*}

The solutions to $\Omega_\chi h^2 = 0.12$ are shown by the horizontal portions of the curves in Fig.~\ref{fig:test1}.
In this regime, the upper limit in the reheating temperature (to ensure that $\Omega_\chi h^2 \le 0.12$) from
Eq.~(\ref{Eq:omegaIA}) is  
$\trh \lesssim 7.7 \times 10^{-{23}}m_\chi^2$
(all quantities in GeV). Thus, even a spectator mass equal to the inflaton mass would require a reheating temperature $\lesssim 69$~TeV, using $H_{\rm end}$ given in Eq.~(\ref{summ}). This is consistent with the results found in \cite{Cosme:2020nac}. Note also the constraint $\trh \gtrsim 4$ MeV to ensure successful nucleosynthesis \cite{tr4} imposes
a {\it lower} bound on $m_\chi \gtrsim 7\times 10^{9}$ GeV.

In addition to the contribution to the relic density from the oscillations of $\chi$, there are two additional sources of $\chi$ production. One is the single graviton exchange process involving
particle production from the inflaton condensate. The other is the four-point interaction. However, for small $\tilde \sigma$, the relic density from long wavelength far exceeds that obtained from particle production through single graviton exchange. 

The rate per unit volume for gravitational production \cite{cmov,Clery:2022wib,Mambrini:2022uol} including the four-point interaction is given by 
\begin{align}
R_g= &\frac{ 2 \times \rho_\phi^2}{256 \pi M_P^4}
\left(1-4\tilde{\sigma}+\frac{m_{\chi,{\rm eff}}^2}{2m^2_\phi}\right)^2\sqrt{1-\frac{m_{\chi,{\rm eff}}^2}{m_\phi^2}} 
\, ,
\end{align}
where the explicit factor of 2 indicates that two $\chi$ particles are produced per interaction. To obtain the relic density, this rate should be integrated in the Boltzmann equation,
\beq
\frac{dY_\chi}{da}=\frac{\sqrt{3}M_P}{\sqrt{\alpha} \trh^2}a^2\left(\frac{a}{a_{\rm RH}}\right)^{\frac32}R_g(a) \, ,
\label{Eq:boltzmann4}
\eeq
where the energy density of $\chi$ is $\rho_\chi = m_{\chi,{\rm eff}} Y_\chi/a^3$.
When $m_{\chi,{\rm eff}}^2 \ll m_\phi^2$ (and ${\tilde \sigma} \ll 1$), there is a simple analytic solution \cite{cmov}, and the relic density can be approximated by
\beq
\frac{\Omega_\chi h^2}{0.12} \simeq 8.6 \times 10^6 \frac{H_{\rm end} \trh m_\chi}{{\rm GeV} M_P^2} \, .
\label{cmov}
\eeq

This becomes comparable to the density given in Eq.~(\ref{Eq:omegaIA}) only when $m_\chi \simeq 1.1 \times  10^{13}$~GeV. 
Note that the relic density from single graviton exchange is proportional to $m_\chi$, whereas the density from large-scale fluctuations is inversely proportional to the mass squared. Combining both sources of production, we see that there is an additional an upper limit to $\trh \lesssim 4.3$~TeV.
Thus, in the absence of even a small (Planck-suppressed) coupling of the spectator to the inflaton, the reheating temperature
is restricted to a range $0.004 < \trh/{\rm GeV} < 4300$, and the mass of the spectator is then restricted to the range $ 7 \times 10^{9} < m_\chi/{\rm GeV} <  3 \times 10^{13}$ (the upper limit corresponding to the inflaton mass and hence kinematic limit).  As we will see,
the gravitational strength coupling, when included, greatly opens up the allowed range for the reheating temperature and spectator mass.

The spectator field may begin oscillating after reheating ($a_{\rm osc}> a_{\rm RH}$ corresponding to $3 \alpha \trh^4 > m_\chi^2 M_P^2$). We refer to this as case IB. In this case, we can evolve $\rho_\chi$ directly from $a_{\rm osc}$ to present time using 
\begin{equation}
\begin{aligned}
  & \rho_{\chi}(a_0) \; =  \; \rho_\chi(a_{\rm osc}) \left(\frac{a_{\rm osc}}{a_0} \right)^3 \\ 
  & \; = \; \frac{3} {16\pi^2} H_{\rm end}^4 
  \frac{m_\chi^2}{m_\chi^2 + \sigma \phi_{\rm end}^2} 
  \left( \frac{g_0}{g_{\rm osc}} \right)
  \left(\frac{T_0}{T_{\rm osc} }\right)^3  \\ & \; \simeq \;    \frac{3}{16\pi^2} H_{\rm end}^4  \frac{(3\alpha)^\frac34  T_0^3}{2^\frac32 M_P^\frac32 m_\chi^\frac32}
  \left(\frac{g_0}{g_{\rm osc}}\right)\, ,
  \label{rhochiarhb}
\end{aligned}
\end{equation}
when $a_{\rm \chi}  < a_{\rm RH} < a_{\rm osc}$ and is independent of the reheating temperature. To obtain the last line of Eq.~(\ref{rhochiarhb}),
we used $m_{\chi,{\rm eff}}^2(a_{\rm osc}) \simeq m_\chi^2 = \frac94 H^2 = \frac34 \alpha T_{\rm osc}^4/M_P^2$.

The present fraction of critical density is then given by Eq.~(\ref{eq:dmdensity})
\begin{equation}
\begin{aligned}
& \frac{\Omega_{\chi} h^2}{0.12} \; = \; 1.1 \times 10^{7} {\rm GeV}^{-1} \frac{H_{\rm end}^4} {M_P^{\frac32} m_\chi^{\frac32} } 
    \\
& \simeq \left(\frac{4.3 \times 10^{20}~{\rm GeV}}{m_\chi}\right)^\frac32\, ,~~~~\sigma \phi_{\rm end}^2 \ll m_\chi^2 \,.
\label{rhochiaosc2}
\end{aligned}
\end{equation}
It is then clear that the Universe is largely over-closed
in this case, as the condition $\Omega h^2 \lesssim 0.12$ requires $m_\chi > M_P$. We do not consider this case any further.

\subsection{Case II: $\sigma \phi_{\rm end}^2 \gg m_{\chi}^2$}

As $\sigma$ is increased, the dependence on $m_\chi$ in Eq.~(\ref{rhochiarh}) disappears (as long as $m_\chi^2 \ll \sigma \phi_{\rm end}^2$) and $\trh$ increases. As long as ${\tilde \sigma} < 9/32$, we maintain, $\aend < a_\chi < a_{\rm osc}$ and the 
relic density (\ref{Eq:omegaIA}) becomes
\begin{equation}
    \frac{\Omega_\chi h^2}{0.12} \; \simeq \; 2.5 \times 10^7 \, \mathrm{GeV}^{-1} \frac{H_{\rm end}^4 T_{\rm RH}}{M_P^2 \sigma \phi_{\rm end}^2  } 
    \simeq\tilde \sigma^{-1}\frac{\trh}{25~{\rm TeV}}
    \, .
    \label{Eq:omegaIAbis}
\end{equation}
Thus for a fixed relic density, we require an increase in $\trh \sim 25 {\tilde \sigma}$~GeV as seen in Fig.~\ref{fig:test1}.
The curves depart from being horizontal at low ${\tilde \sigma}$ when ${\tilde \sigma} \simeq m_\chi^2/4 H_{\rm end}^2$. They then appear to merge (due to the log scale of the plot) as $\trh$ increases with $\tilde \sigma$ almost independently of $m_\chi$.
This behavior is very similar to Eq.~(\ref{Eq:omegaIA}), with
$\tilde \sigma H_{\rm end}^2$ playing the role of $m_\chi^2$.

For ${\tilde \sigma} > 9/32$, the spectator field begins to oscillate while dominated by the effective mass $\propto \phi$, until $a=\achi$,
where the oscillations are dominated by the bare mass.
Thus, $\aosc<\achi$ (see Eq.~(\ref{achiltaosc})). For lower values of $\tilde \sigma$, $a_\chi < a_{\rm osc}$, and case I is applicable. 
Note that, as $\sigma \phi^2$ redshifts as $a^{-3}$, as does $H^2$, 
once the oscillations of $\chi$ begin ($m_{\chi,{\rm eff}}> \frac32 H$) they never stop .
Indeed, the effective
mass term $\propto \phi$ decreases with time, and the bare mass
dominates the oscillations at $a=\achi$.
It is easy to understand that for
larger $\tilde \sigma$ the phase of oscillations begin earlier because $\sigma \phi^2 > m_\chi$. 
During this phase of oscillations, $\rho_\chi$ undergoes a pre-dilution 
between $\aend$ and $\achi$, opening the $\trh$ parameter space
relative to that allowed when only the bare mass term is present.

Similarly to case IA, we will consider first the late reheating scenario, 
{\it i.e.}, $a_{\chi} < \arh $, or $\sigma \phi^2(\arh)\lesssim m_\chi^2$, with the upper limit on $\tilde \sigma$ given by Eq.~(\ref{acltarh}).
We refer to this as case IIA, which allows for much larger $\trh$ than case IA.

The energy density must now be evolved from $a_{\rm end}$ as follows:
\begin{equation}
\begin{aligned}
 \rho_\chi(a_0) \; = \;  & \frac{3\beta}{16\pi^2} H_{\rm end}^4 \left( \frac{a_{\rm end}}{a_{\rm osc}} \right)^3 \left( \frac{a_{\rm osc}}{a_{\chi}} \right)^{3+2A} 
 \\ &  \times   \left( \frac{a_{\chi}}{a_{\rm RH}} \right)^3 \left( \frac{a_{\rm RH}}{a_0} \right)^3  \\ = & \frac{3}{16\pi^2} H_{\rm end}^4 \left( \frac{a_{\rm osc}}{a_{\rm end}} \right)^{2A} \left( \frac{a_{\rm end}}{a_\chi} \right)^{2A} \\ 
 & \; \times \; \left( \frac{a_{\rm end}}{a_{\rm RH}} \right)^3  \left( \frac{a_{\rm RH}}{a_0} \right)^3 \, ,
 \label{rhochiIIA}
\end{aligned}
\end{equation}
where $A$ is defined from Eq.~(\ref{eq:chi(a)})
 For $\tilde \sigma = \frac{9}{32}$, using the relations in Eqs.~(\ref{aoae})-(\ref{arhae}), this solution matches that given in Eq.~(\ref{eq:dmdensity}).
For the limited range $9/32 < {\tilde \sigma} < 9/16$, $\aend < a_{\rm osc}$.  For larger ${\tilde \sigma} > 9/16$,
$a_{\rm osc} = \aend$.
However, in this case,  the exponential suppression in the initial value of $\langle \chi^2 \rangle$ alluded to earlier,
\beq
\beta = {\rm Min} \left(\exp \left[{-1.25 \pi \left(4 {\tilde \sigma} +\frac{m_\chi^2}{H_{\rm end}^2} - \frac94 \right)}\right], 1 \right) \, ,
\label{eq:expCASEIIA}
\eeq 
must be included.

\begin{figure}[t!]
  \centering
\includegraphics[width=0.48\textwidth]{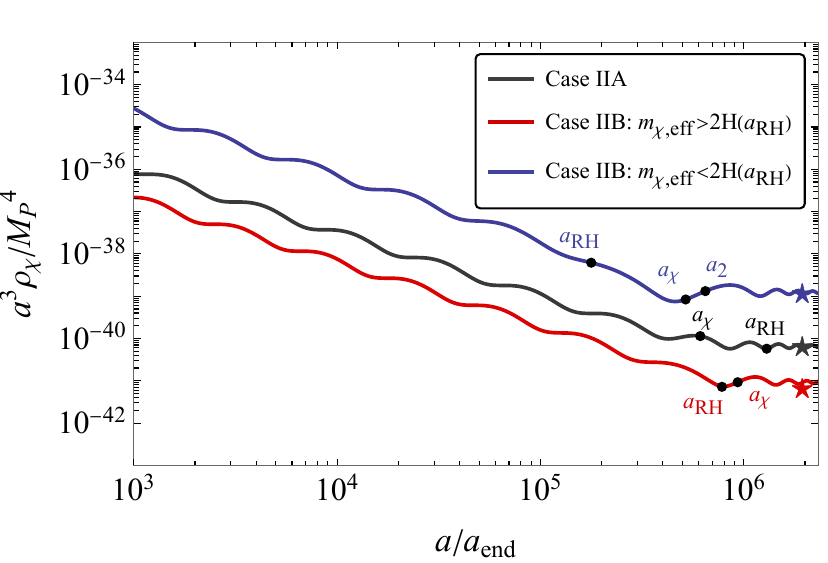}
  \caption{\em \small {
  The evolution of the energy density of the spectator field $\rho_\chi a^3$ for case IIA (black), case IIB with $m_{\chi,{\rm eff}} > 2 H(\arh)$ (red), and case IIB with $m_{\chi,{\rm eff}} < 2 H(\arh)$ (blue).
$a_2$ denotes $a_{\text{osc-2}}$}. 
Parameters chosen are:  $\tilde \sigma = 1.7$, $m_\chi = 5\times 10^4$~GeV, and $\trh = 5.95 \times 10^{10}$~GeV for the black curve; $\tilde{\sigma}= 1.8$, $m_\chi = 3\times 10^4$~GeV, and $\trh = 8.7\times 10^{10}$~GeV for the red curve; and $\tilde{\sigma}= 1.5$, $m_\chi =  1.5\times 10^4$~GeV, and $\trh = 2.64\times 10^{11}$~GeV (blue curve). In each case, the star denotes the present day relic density from Eqs.~(\ref{eq:caseIIAOmegachi}), (\ref{IIBi}), and (\ref{case2B2relic}), respectively. In all three cases, $\aosc = \aend$. The parameters in each case are chosen so that $\Omega_\chi h^2=0.12$ when both production mechanisms are considered. 
}
  \label{fig:rhochi2AB}
\end{figure}

In this case, the resulting density fraction is
\begin{equation}
    \Omega_{\chi} h^2 \; = \; 6.7 \times 10^5~\beta~\frac{H_{\rm end} m_{\chi} T_{\rm RH}}{\mathrm{GeV} M_P^2 \sqrt{\tilde \sigma} } \, . \label{eq:caseIIAOmegachi}
\end{equation}
or
\beq
\frac{\Omega_\chi h^2}{0.12}\simeq \frac{\beta}{\sqrt{\tilde \sigma}}
\left(\frac{\trh}{10^{10}~{\rm GeV}}\right)
\left(\frac{m_\chi}{1.4\times 10^7~{\rm GeV}}\right)\,.
\label{eq:caseIIAOmegachi2}
\eeq
The numerical solution to the equation of motion for $\chi$ in case IIA with ${\tilde \sigma} = 1.7$ is shown in Fig.~\ref{fig:rhochi2AB}. For this value of $\tilde \sigma$, $a_{\rm osc} = \aend$, $a_\chi \simeq 6.1\times 10^5 \aend$, and $a_\text{RH} \simeq 1.3 \times 10^6 \aend$. In contrast to the case shown in Fig.~\ref{fig:enden1},
oscillations begin when inflation ends and the density decreases as $a^{-\frac92}$. For $a > a_\chi$, the density drops as $a^{-3}$. Once again, we see excellent agreement between the numerical and analytical results.

As one can discern from Eq.~(\ref{eq:caseIIAOmegachi2}), the allowed range for $m_\chi$ and $\trh$ is considerably greater than that found for case IA, when the coupling $\sigma$ could be ignored. In this case, we also see that the relic density is proportional to $m_\chi \trh$, whereas at small ${\tilde \sigma}$, the relic density of $\chi$ was dominated by the large-scale fluctuations, as ${\tilde \sigma}$ is increased, the contribution from particle production (either from single graviton exchange or directly from the contact term) comes to dominate. Recall that for small $\tilde \sigma$, the two contributions become comparable only at large bare mass $m_\chi \simeq 10^{13}$~GeV. 
When $\tilde \sigma > 9/16$, the density produced from single graviton exchange and the contact interaction begin to dominate for all masses. Indeed, for $\tilde \sigma > 9/16$ there are two sources of suppression of the large wavelength modes. One is the exponential suppression when the effective mass is large and this occurs when ${\tilde \sigma} = 9/16 - m_\chi^2/4 H_{\rm end}^2$. However, as noted earlier, when 
${\tilde \sigma} = 9/16$, $\aosc = \aend$, and we lose a strong enhancement of $(\aosc/\aend)^{3/2} \gg 1$ at smaller $\tilde \sigma$.

These effects are displayed in both panels of Fig.~\ref{fig:test1}. 
At $\tilde \sigma = 9/16$, we see a meteoric rise in the required reheating temperature to attain $\Omega_\chi h^2 = 0.12$. The degree to which the temperature rises depends on the bare mass and is roughly inversely proportional to $m_\chi$. At large $\tilde \sigma$, the relic density is predominantly due to particle production and obtained from a numerical integration of Eq.~(\ref{Eq:boltzmann4}). Because $m_{\chi,{\rm eff}}^2 = m_\chi^2 + \sigma \phi^2(a)$ is a function of the scale factor $a$, a {\em useful} analytical form for $\rho_\chi$ is not available.  Furthermore, as ${\tilde \sigma}$ is increased, the effective mass of $\chi$ increases, and eventually the particle production processes are kinematically forbidden when inflation ends.  However, because the effective mass depends on $a$, as $a$ increases, particle production is again kinematically allowed. This effect first occurs when $\tilde \sigma \simeq 2.5$ and is visible in the plot as a small change in slope at $\log {\tilde \sigma} \simeq 0.4$.

We now see the full effect of adding the interaction term for $\chi$. In its absence, we are restricted to very large scalar masses ($10^{10} - 10^{13}$)~GeV and very low reheating temperatures ($0.004 - 4300)$~GeV as seen in the horizontal portions of the curves in the left panel of Fig.~\ref{fig:test1}. For non-negligible ${\tilde \sigma} \le 9/16$, the reheating temperature may be as large as the allowed maximum $= (3/\alpha)^\frac14 (H_I M_P)^\frac12 = 2.3\times 10^{15}$~GeV, where the maximal value is attained for low masses $m_{\chi} \simeq 50$~GeV. At ${\tilde \sigma} = 9/16$, the contribution of the large-scale fluctuations is highly suppressed and the relic density is produced by particle production from the inflaton condensate. The inclusion of a coupling between the inflaton and the spectator now allows for a wide range of masses and reheating temperatures all satisfying $\Omega_\chi h^2 = 0.12$.

\begin{figure}[t!]
  \centering
\includegraphics[width=0.48\textwidth]{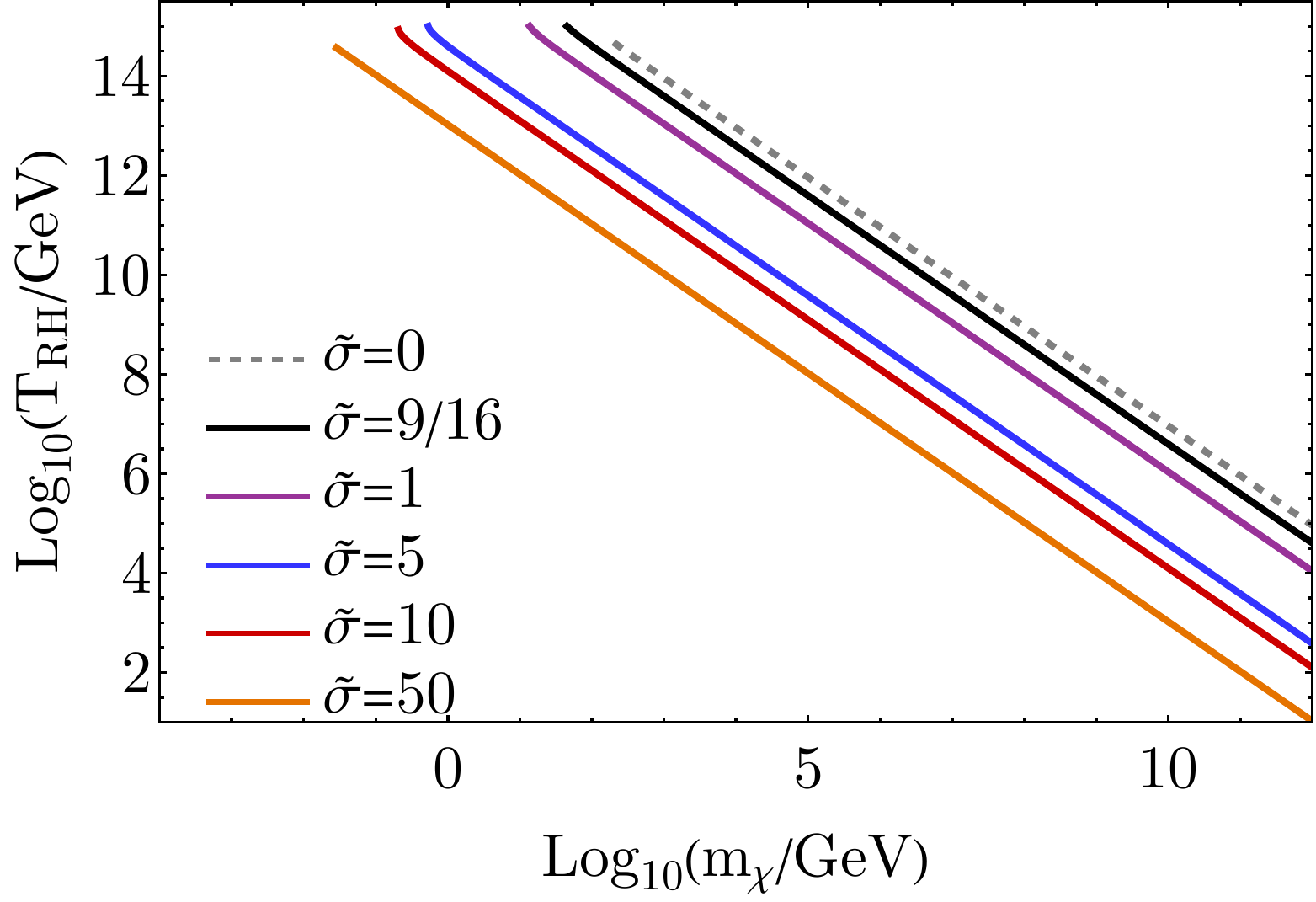}
  \caption{\em \small {The reheating temperature, $T_{\rm RH}$ vs the bare mass, $m_\chi$, satisfying $\Omega_{\chi}h^{2}=0.12$ for $\tilde{\sigma} = \frac{9}{16}$(black), 1 (purple),  $5 $(blue), $10$(red), $50$(orange), $0$(gray dashed). The line  for $\tilde{\sigma}=0$ neglects the contribution from large-scale fluctuations and corresponds to the result in \cite{cmov}.  For $m_{\chi}<m_{\rm \chi,min}$, (corresponding to the endpoint of the lines) there is no $T_{\rm RH}$ satisfying $\Omega_{\chi}h^{2}=0.12$.
}
}
  \label{fig:test2}
\end{figure}

In Fig.~\ref{fig:test2}, we show the allowed parameter space for ${\tilde \sigma} \ge 9/16$ in the $(\trh,m_\chi)$ plane. The curves correspond to $\Omega_\chi h^2 = 0.12$ for fixed values of $\tilde \sigma$ as indicated.  As noted above, for ${\tilde \sigma} \ge 9/16$, the relic density is saturated by the contribution from inflaton scattering. Here, we see that the temperature is maximal at low masses and that this maxiumum occurs at lower $m_\chi$ as $\tilde \sigma$ is increased. The dotted line in  Fig.~\ref{fig:test2} shows the result from \cite{cmov} due to inflaton scattering through single graviton exchange (i.e., with $\tilde \sigma = 0$) neglecting the (dominant) contribution from large scale fluctuations. In this case there is direct relation between $m_\chi$ and $\trh$ as given in Eq.~(\ref{cmov}). For masses lower than $\sim 100$~GeV, there are no solutions that yield the correct relic density (smaller masses and/or reheating temperatures give $\Omega_\chi h^2 < 0.12$).
As $\tilde \sigma$ is increased, the lower limit on the mass decreases. For example, for ${\tilde \sigma} = 1$, $m_\chi \gtrsim 35$~GeV, the maximal temperature cannot be obtained as kinematic blocking due to the large effective mass prevents particle production, and so the reheating temperature must be lower so that $\arh$ is large enough to allow a finite range of integration in Eq.~(\ref{Eq:boltzmann4}). In this case,  $m_\chi \gtrsim 10$~GeV and the maximum reheating temperature is $\trh \sim 1.1 \times 10^{15}$. For ${\tilde \sigma} = 50$,  $m_\chi> 0.03$~GeV and $\trh \simeq 8\times10^{14}$~GeV is required to obtain $\Omega_\chi h^2 = 0.12$. 

Included in the solutions shown in Figs.~\ref{fig:test1} and \ref{fig:test2}, are solutions corresponding to case IIA as well as other evolutionary patterns. For completeness, we describe these possibilities. 
When ${\tilde \sigma} > \frac{9}{32}$ and ${\tilde \sigma} > 3 m_\chi^2 M_P^2/4 \alpha \trh^4$, $a_{\rm osc} < a_{\rm RH} < a_{\chi} $ and we refer to this as case IIB. As before, for ${\tilde \sigma} > \frac{9}{16}$, $a_{\rm osc} = a_{\rm end}$ (for smaller ${\tilde \sigma}$ there is an additional suppression that can be obtained from Eq.~(\ref{aoae})). 
However, when $a_\chi > a_{\rm RH}$, the effective mass after reheating quickly becomes equal to the bare mass as the inflaton decays. The spectator density can then be computed using $m_\chi^2 \langle \chi^2 \rangle$ and allowing $\chi$ to evolve as $a^{-2A}$ and providing a factor $(\aend/a_{\rm RH})^{3/2}$. After reheating (or $a > a_\chi$ which is only slightly larger than $a_{\rm RH}$) the density falls as $a^{-3}$ providing the factor in Eq.~(\ref{today}). Putting all the factors together, we arrive at
\begin{equation}
      \frac{\Omega_{\chi} h^2}{0.12} \; = \; \frac{3.2 \times 10^{6} H_{\rm end}^3 m_{\chi}^2 \beta}{\mathrm{GeV} M_P T_{\rm RH} (m_\chi^2 + \sigma \phi_{\rm end}^2)} \, .
      \label{IIBi}
\end{equation}

The evolution of $\rho_\chi$ for this case with $a_{\rm osc} < a_{\rm RH} < a_{\chi} $ is shown by the red curve in Fig.~\ref{fig:rhochi2AB}.
Note that due to inflaton decay, the value of $a_\chi$ is slightly larger than that obtained from Eq.~(\ref{acae}), though this is not used in the numerical integration. The analytic result from Eq.~(\ref{IIBi}) gives very good agreement with the numerical result.

In this case, we must further distinguish an additional possibility. When $m_{\chi,{\rm eff}} > 2H(a_{\rm RH})$ oscillations of $\chi$ continue uninterrupted.\footnote{As the Universe is radiation-dominated, the condition for oscillations is $m = 2H$ rather than $m=3H/2$ as used earlier in the matter-dominated era.}
This led to the solution given in Eq.~(\ref{IIBi}).
If, however, $m_{\chi,{\rm eff}}$ is small and $m_{\chi,{\rm eff}} < 2H(a_{\rm RH})$ after reheating, there will be a delay in the oscillations of $\chi$, until $a_{\rm osc-2}$ when $m_{\chi,{\rm eff}} = 2H(a_{\rm osc-2})$. Relative to the previous result in Eq.~(\ref{IIBi}), there is a slight enhancement by a factor $(a_{\rm osc-2}/a_{\rm RH})^3 = (\trh/T_{\rm osc-2})^3$, where $T_{\rm osc-2}$ is given by a similar expression used in Eq.~(\ref{rhochiarhb}), $4 \alpha T_{\rm osc-2}^4 = 3 m_\chi^2 M_P^2$.  The result is
\begin{equation}
     \frac{\Omega_{\chi} h^2}{0.12} \; = \; \frac{5.7 \times 10^{7} H_{\rm end}^3 \sqrt{m_{\chi}} T_{\rm RH}^2 \beta}{\mathrm{GeV} M_P^{5/2} (m_\chi^2 + \sigma \phi_{\rm end}^2)} \, .
     \label{case2B2relic}
\end{equation}
The evolution of $\rho_\chi$ for this case is shown by the blue curve in Fig.~\ref{fig:rhochi2AB}. Note that for $a_{\rm osc-2} > a > a_\chi$, the evolution of $\chi$ stops and $\rho_\chi$ is constant, leading to the observed increase in $a^3 \rho_\chi$ seen in the figure. Once again, for this parameter set, the agreement between the numerical evaluation and the analytical result is excellent.

\section{Discussion}
\label{sec:discussion}

In this work, we have considered the production of a massive (stable) scalar field during inflation and the subsequent reheating process. It is well known \cite{fluc} that for scalar 
fields with masses significantly below the Hubble scale during inflation, large-scale fluctuations are produced with $\langle \chi^2 \rangle \gg H_I^2$ as given in Eq.~(\ref{eq:exptvalS}). 
When inflation ends, these fluctuations can be viewed as long-wavelength oscillations, and their contribution to the energy 
density takes the form of matter, which can account for all or part of the dark matter. Indeed, for light fields, this form of 
production is so efficient that there are strong constraints on the mass and reheating temperature to avoid the overproduction of dark matter \cite{Turner:1987vd,Peebles:1999fz,Enqvist:2014zqa,Nurmi:2015ema,Bertolami:2016ywc,Alonso-Alvarez:2018tus,Markkanen:2018gcw,Tenkanen:2019aij,Cosme:2020nac,Lebedev:2022cic}, see Eq.~(\ref{Eq:omegaIA}). This can be seen graphically in 
Fig.~\ref{fig:test1} in the limit $\tilde \sigma\rightarrow 0$, which displays curves of constant $\Omega_\chi h^2 = 0.12$ for fixed $m_\chi$.

In addition, after inflation, there is the inevitable production of scalars through their minimal coupling to 
gravity \cite{MO,cmov}. This source of production can be 
understood as a single graviton exchange coupling the inflaton 
condensate to the scalar. In the absence of scalar self-interactions or direct coupling to the inflaton the energy density resulting from the large-scale fluctuations dominate the total.

In this work, we have considered the possibility of a Planck-suppressed coupling between the scalar and the inflaton. Indeed, such couplings are generated by graviton exchange, but these are derivative couplings that do not affect the effective mass of $\chi$. Nevertheless, without a symmetry preventing such a coupling, we would expect Planck-suppressed coupling, $\sigma \phi^2 \chi^2$, to be present. Such a coupling will also be generated if $\chi$ is non-minimally coupled to curvature \cite{Clery:2022wib}. We have normalized the coupling to the inflaton mass
so that $\sigma = {\tilde \sigma} m_\phi^2/M_P^2$. 

For extremely small $\tilde \sigma$, we have seen that the relic density is indeed determined almost solely by the large-scale fluctuations and therefore there are very strong constraints on the mass and reheating temperature due to the overproduction of $\chi$ during the phase of inflation. These are summarized in Eq.~(\ref{Eq:omegaIA}). 
However, even for very small coupling, $\tilde{ \sigma} \gtrsim 10^{-7}$,
the allowed reheating temperature range is increased as sensitivity to the scalar mass is decreased, as can be seen in Fig.~\ref{fig:test1}. This is due to an earlier phase of oscillations initiated by the $\sigma \phi^2$ term and enhanced dilution.

For larger ${\tilde \sigma} \ge \frac{9}{16}$, direct particle production from the condensate becomes important and soon dominates. In this regime, there is a wide range of scalar masses (from 50 MeV to $10^{13}$~GeV) and a wide range in reheating temperatures (from $\sim 1$ to $\sim 10^{15}$~GeV) for $0.004 < \tilde{\sigma}  < 50$, where the lower bound comes from the constraint obtained from isocurvature perturbations and the upper bound from ensuring that non-perturbative processes such as parametric resonance is unimportant. These are our main results and are displayed in the right panel of Fig.~\ref{fig:test1} and Fig.~\ref{fig:test2}.

We have provided a combination of analytical and numerical results to calculate the relic density. A more complete numerical analysis will be presented elsewhere \cite{paper2}. These results can naturally be generalized by including scalar self-interactions $\lambda_p \chi^p$, as well as considering background cosmologies in which the inflaton potential is not quadratic about its minimum. The case for $p = 4$ was considered in \cite{Markkanen:2018gcw,Cosme:2020nac}.

\acknowledgements
This project has received support from the European Union's Horizon 2020 research and innovation program under the Marie Sklodowska-Curie grant agreement No 860881-HIDDeN and the CNRS-IRP project UCMN. The authors would like to acknowledge Simon Clery, Mathieu Gross, and Jong-Hyun Yoon for useful discussions. The work of G.C. and K.A.O.~was supported in part by DOE grant DE-SC0011842 at the University of Minnesota. The work of M.A.G.G.~was supported by the DGAPA-PAPIIT grant IA103123 at UNAM, the CONAHCYT ``Ciencia de Frontera'' grant CF-2023-I-17, and the Programa de Investigaci\'on del Instituto de F\'isica 2023 (PIIF23). S.V. was supported in part by DOE grant DE-SC0022148 at the University of Florida.

\clearpage

\onecolumngrid

\appendix

\section{Fokker-Planck Approach}
\label{app:fp}
To compute the expectation value of $\chi$, we use the stochastic approach that treats the long wavelength superhorizon contribution of the field $\chi(t, \mathbf{x})$ in a de Sitter background as a classical stochastic variable, $\varphi$, characterized by the probability distribution $\rho(t, \varphi)$~\cite{Starobinsky:1986fx, Starobinsky:1994bd, Kamenshchik:2021tjh}. This probability distribution must satisfy the Fokker-Planck equation
\begin{equation}
    \frac{\partial \rho}{\partial t} \; = \; \frac{H_I^3}{8\pi^2} \frac{\partial^2 \rho}{\partial \varphi^2} + \frac{1}{3H_I} \frac{\partial}{\partial \phi} \left(\frac{\partial V}{\partial \varphi} \rho(t, \varphi) \right) \, .
\end{equation}
This expression implies that the long wavelength contribution of an effective potential $V(s(t, \mathbf{x}))$ is equal to the expectation value of the effective potential of the classical stochastic variable $V(\varphi)$. Therefore, using the effective mass, $m_{\chi, \rm eff}$, the spectator field potential in terms of the classical stochastic variable can be expressed as
\begin{equation}
    \label{eq:effpotquartic}
    V(\varphi) \; = \; \frac{1}{2}m_{\chi, \rm eff}^2 \varphi^2 \, .
\end{equation}
At late times, the solution to the Fokker-Planck equation reaches static equilibrium and becomes
\begin{equation}
\rho_{\mathrm{eq}}(\varphi)=N^{-1} \exp \left(-\frac{8 \pi^2}{3 H_I^4} V(\varphi)\right) \, ,
\end{equation}
where $N$ is fixed according to the normalization condition
\begin{equation}
    \int_{-\infty}^{\infty} \rho_{\mathrm{eq}}(\varphi) d \varphi=1 \, .
\end{equation}
For the effective potential~(\ref{eq:effpotquartic}), we find that the normalization constant $N$ can be expressed as
\begin{equation}
N \; = \; \int_{-\infty}^{\infty} \exp \left[-\frac{8 \pi^2}{3 H_I^4}\left(\frac{m_{\chi, \rm eff}^2 \varphi^2}{2}\right)\right] d \varphi \; = \; \sqrt{\frac{3H_I^4}{4\pi m_{\chi, \rm eff}^2}} \, .
\end{equation}
Next, we compute the expectation value of $\langle \chi^2 \rangle$, given by
\begin{equation}
    \begin{aligned}
    \langle \chi^2 \rangle \; = \; \langle \varphi^2 \rangle \; = \; \int_{-\infty}^{\infty} \varphi^2 \rho_{\rm eq}(\varphi) d\varphi \; = \; \frac{1}{N} \frac{3\sqrt{3} H_I^6}{16 \pi^{\frac{5}{2}} m_{\chi, \rm eff}^3} \; = \; \frac{3H_I^4}{8 \pi^2 m_{\chi, \rm eff}^2}\, .
    \end{aligned}
\end{equation}
We note that when considering a pure de Sitter background, the asymptotic value is determined by the Hubble parameter, $H_I$. However, numerical studies show that for a slow-roll model of inflation, the density of $\chi$ is a better fit using $H_{\rm end}$ \cite{paper2}. Thus, at the end of inflation, we use the following expectation value for the spectator field:
\begin{equation}
    \langle \chi^2 \rangle_{\rm end} \; \simeq \frac{3H_{\rm end}^4}{8 \pi^2 m^2_{\chi, \rm eff}(a_{\rm end})} \, .
\end{equation}

\end{document}